\documentclass{article}

\usepackage{PRIMEarxiv}
\DeclareUnicodeCharacter{202F}{\,}
\usepackage[utf8]{inputenc} 
\usepackage[T1]{fontenc}    
\usepackage{hyperref}       

\usepackage{url}            
\usepackage{booktabs}       
\usepackage{amsfonts}       
\usepackage{nicefrac}       
\usepackage{microtype}      
\usepackage{lipsum}
\usepackage{fancyhdr}       
\usepackage{graphicx}       
\graphicspath{{media/}}     
\usepackage{amsmath}
\usepackage{amssymb}
\usepackage{amsfonts}

\usepackage[table]{xcolor}

\usepackage{graphicx}
\usepackage{subcaption}

\usepackage{url}
\usepackage{microtype}
\usepackage{nicefrac}
\usepackage{lipsum} 
\usepackage{blindtext} 

\usepackage{booktabs}
\usepackage{longtable}
\usepackage{multirow}
\usepackage{diagbox}
\usepackage{tabularx}
\usepackage{threeparttable}

\usepackage{algorithm}
\usepackage{algpseudocode}

\usepackage[section]{placeins}

\usepackage[utf8]{inputenc}
\title{Automating Crash Diagram Generation Using Vision-Language Models: A Case Study on Multi-Lane Roundabouts
}

\author{
  Xiao Lu, Hao Zhen, Jidong J. Yang*\\
  Smart Mobility and Infrastructure Lab, College of Engineering \\
  University of Georgia \\
  Athens\\
  \texttt{\{Xiao.Lu1, Hao.Zhen, Jidong.Yang\}@uga.edu} \\
}

\begin{document}

\maketitle

\begin{abstract}
Crash diagrams are essential tools in transportation safety analysis, yet their manual preparation remains time-consuming and prone to human variability. This study investigates the use of Vision-Language Models (VLMs) to automate crash diagram generation from police crash reports, focusing on multilane roundabouts as a challenging test case. A three-part structured prompt framework was developed to guide model reasoning through interpretation, extraction, and visual synthesis, while a 10-metric evaluation system was designed to assess diagram quality in terms of semantic accuracy, spatial fidelity, and visual clarity. Three popular models, including GPT-4o, Gemini-1.5-Flash, and Janus-4o, were tested on 79 crash reports. GPT-4o achieved the highest average performance (6.29 out of 10), followed by Gemini-1.5-Flash (5.28) and Janus-4o (3.64). The analysis revealed GPT-4o’s superior spatial reasoning and alignment between extracted and visualized crash data. These results highlight both the promise and current limitations of VLMs in engineering visualization tasks. The study lays the groundwork for integrating generative AI into crash analysis workflows to improve efficiency, consistency, and interpretability.
\end{abstract}


\section{Introduction}

Crash diagrams are essential tools in transportation safety analysis, providing standardized visual representations of collision configurations, vehicle trajectories, and contributing factors at crash locations. These diagrams serve as fundamental components of state and national safety databases, enabling transportation agencies to identify crash patterns, recognize systemic hazards, and develop targeted safety countermeasures~\cite{fernandez2025avoiding}. The visual nature of crash diagrams facilitates rapid comprehension of complex collision dynamics, making them indispensable for safety engineers, law enforcement, and policymakers in their efforts to reduce traffic-related injuries and fatalities.

Despite their critical importance, the current process of generating crash diagrams remains largely manual and resource-intensive. Trained traffic safety professionals must interpret written crash narratives, which are often fragmented, ambiguous, or incomplete, and translate them into standardized visual representations. This manual process presents several significant challenges: it requires substantial time investment, with each diagram taking 15--30 minutes to create;  it demands specialized expertise in crash reconstruction and diagram conventions; it is prone to interpretation variability between different analysts; and it creates scalability bottlenecks when processing large volumes of crash data~\cite{jaradat2025ai}. These limitations are particularly pronounced for geometrically complex intersections such as multilane roundabouts, where the circular geometry, multiple conflict points, and intricate vehicle paths demand heightened spatial reasoning capabilities. {In a multilane roundabout, vehicles enter from multiple approaches, circulate within two concentric lanes around a central island, and exit at downstream legs, resulting in curved trajectories, lane-changing maneuvers, and multiple conflict locations at entries, within the circulatory roadway, and at exits. This geometric complexity increases the difficulty of accurately interpreting and depicting crash scenarios.}

In current practice, beside the manual crash diagram drawing, agencies typically rely on established software tools to assist with collision visualization and crash pattern identification. Examples include the crash diagram functionality within the Transportation Injury Mapping System (TIMS), which generates collision diagrams from coded crash databases \cite{tims}; commercial platforms such as Crash Magic, which convert structured crash records into collision diagrams and summary analytics \cite{crashmagic}; and AASHTOWare Safety Intersection, building on segment level data, offers intersection-level screening and visualization tools for safety engineers \cite{aashtoware}. These systems demonstrate the longstanding need for streamlined diagram production, yet they also share an important limitation: they depend on pre-coded, structured crash data as input. The interpretation of raw police narratives, damage codes, and officer sketches must still be performed manually before these tools can generate a diagram. {Damage codes in police crash reports are standardized numerical indicators that describe the primary point of impact on each vehicle and provide essential information for reconstructing collision geometry.}

Because existing tools assume that the underlying crash information has already been extracted, standardized, and verified, they do not address the upstream challenge of converting free-text crash reports or scanned forms into complete, spatially coherent diagrams. This gap becomes increasingly significant when processing large volumes of crash reports, particularly for complex geometries such as multilane roundabouts where small interpretive inconsistencies can lead to substantial analytical differences. These limitations highlight the need for an automated approach that begins directly from the original police report, rather than requiring manual preprocessing before visualization is possible.

The advent of generative artificial intelligence presents an unprecedented opportunity to improve the tools at traffic safety. {Vision-Language Models (VLMs) are multimodal artificial intelligence systems that jointly process textual and visual inputs to perform reasoning and generation tasks. In traffic safety applications, police crash reports often include standardized numeric vehicle damage codes that specify the location of impact on each vehicle. A multilane roundabout typically consists of two circulating lanes around a central island, with vehicles entering and exiting through yield-controlled approaches, requiring precise spatial reasoning for accurate crash reconstruction.}
Recent studies have shown the potentials of large language models in traffic crash analysis~\cite{zhen2025tab, zhen2024leveraging, zhen2025crashsage}. Recent breakthroughs in generative VLMs, which can synthesize images from textual descriptions and multimodal inputs, have demonstrated remarkable capabilities in spatial reasoning, scene composition, and visual synthesis~\cite{akter2025large, cao2024maplm}. Unlike earlier AI systems that merely analyzed existing images, modern generative VLMs leverage diffusion-based architectures, transformer-based image synthesis, and cross-modal attention mechanisms to create new visual content that maintains both semantic accuracy and spatial coherence. Furthermore, recent developments in controllable image generation allow these models to produce diagrams that adhere to standardized symbology and formatting conventions used in traffic accident documentation. These capabilities position VLMs as potentially transformative tools for automating the conversion of textual crash reports into standardized visual diagrams. 

However, despite the rapid advancement of VLMs in general image generation tasks, their application to specialized technical domains such as transportation safety remains largely unexplored~\cite{ding2025urban}. The generation of crash diagrams presents unique challenges that distinguish it from generic image synthesis: strict adherence to standardized symbology and diagrammatic conventions; precise spatial localization within constrained geometric environments; accurate interpretation of domain-specific terminology and damage codes; and maintenance of technical accuracy suitable for engineering analysis and legal documentation. To date, no systematic evaluation has been conducted to assess whether state-of-the-art VLMs can meet these stringent requirements. {Rather than proposing a universal diagram-generation methodology, this study focuses on evaluating the feasibility of structured prompting as a diagnostic tool for understanding current vision-language model capabilities in a safety-critical crash diagram synthesis task. Future work will investigate more advanced prompting strategies, such as in-context learning and systematic prompt ablation, to better understand how prompt design influences model reasoning and diagram accuracy.}

This study represents one of the first comprehensive evaluations of generative AI for automated crash diagram generation, establishing both a methodological framework and performance benchmarks for this novel application. We develop a structured three-part prompt engineering approach that systematically guides VLMs through the complex process of interpreting police crash reports, extracting critical information, and synthesizing accurate visual representations. {In this paper, a “structured prompt” refers to a standardized, step-by-step instruction
template that constrains model inputs and outputs to improve consistency across crash cases.
``Ground truth'' refers to the official police crash diagram recorded in the crash report and
used as the reference for evaluation.}
Using multilane roundabouts as a challenging test case—due to their circular geometry, multiple conflict zones, and complex vehicle interactions—we evaluate three leading VLMs: OpenAI's GPT-4o \cite{openai2024chatgpt4o}, Google's Gemini-1.5-Flash \cite{anil2023gemini}, and the open-source Janus-4o \cite{chen2025janus4o}. These models were selected to represent the current spectrum of multimodal AI systems: proprietary (GPT-4o), commercial research (Gemini-1.5-Flash), and open-source (Janus-4o) allowing for a balanced comparison between accessibility, visual quality, and reasoning capability. Unlike prior computer-vision or LiDAR-based approaches that detect real-time accidents through sensor data, this study focuses on post-crash reconstruction from textual police narratives, a domain requiring semantic interpretation and spatial synthesis rather than object detection. This distinction defines the novelty of the work, as it evaluates generative AI’s capacity to perform structured visual reasoning within standardized engineering contexts.

A central contribution of this work is the development of a comprehensive evaluation framework comprising ten specific metrics designed to assess crash diagram quality across multiple dimensions: semantic accuracy (e.g., correct collision type identification), spatial fidelity (e.g., accurate vehicle positioning and impact location), and visual clarity (e.g., proper labeling and proportionality). This metric set, informed by established traffic safety documentation standards, provides the first universal evaluation criteria for AI-generated crash diagrams, establishing a foundation for future research and practical deployment.

Furthermore, we introduce a complete computational pipeline that integrates data preprocessing, prompt construction, model inference, and quality evaluation into a reproducible workflow. This pipeline demonstrates how existing police crash report data can be systematically processed for AI consumption while maintaining the technical details required for safety analysis applications. By evaluating this pipeline on 79 real-world crash cases from a high-volume multilane roundabout in New York State, we provide empirical evidence of both the promise and current limitations of VLMs in this domain.

We carried out the above pipeline for crash diagrams generation of multilane roundabout crashes based on data official New York State Police Accident Reports (Form MV-104A), as part of the ``Reasons for Drivers Failing To Yield at Multi-Lane Roundabout Exits: Transportation Pooled Fund Study.''~\cite{Medina2023} by Federal Highway Administration. Our findings reveal that while state-of-the-art VLMs, particularly GPT-4o, demonstrate significant potential in reducing manual workload and enhancing documentation consistency, important challenges remain in achieving the spatial precision and technical accuracy required for operational deployment. These results not only establish initial performance benchmarks but also identify critical areas for model refinement and dataset expansion. By bridging the gap between cutting-edge AI capabilities and practical transportation safety needs, this study provides actionable insights for researchers, practitioners, and technology developers working toward the responsible integration of artificial intelligence into traffic safety analysis workflows.

The remainder of this paper is structured as follows. First, we introduce the overall framework for automated crash diagram generation and its evaluation approach. Next, we describe the data sources used and the experimental methodology employed to assess model performance. We then present and analyze the results, highlighting the capabilities and limitations of different VLMs along with their implications for transportation safety. Finally, we conclude by summarizing key findings, acknowledging current limitations, and outlining potential directions for future research.

\section{Framework of crash diagram generation and evaluation}
\begin{figure}[htbp!]
    \centering
    \includegraphics[width=1\linewidth]{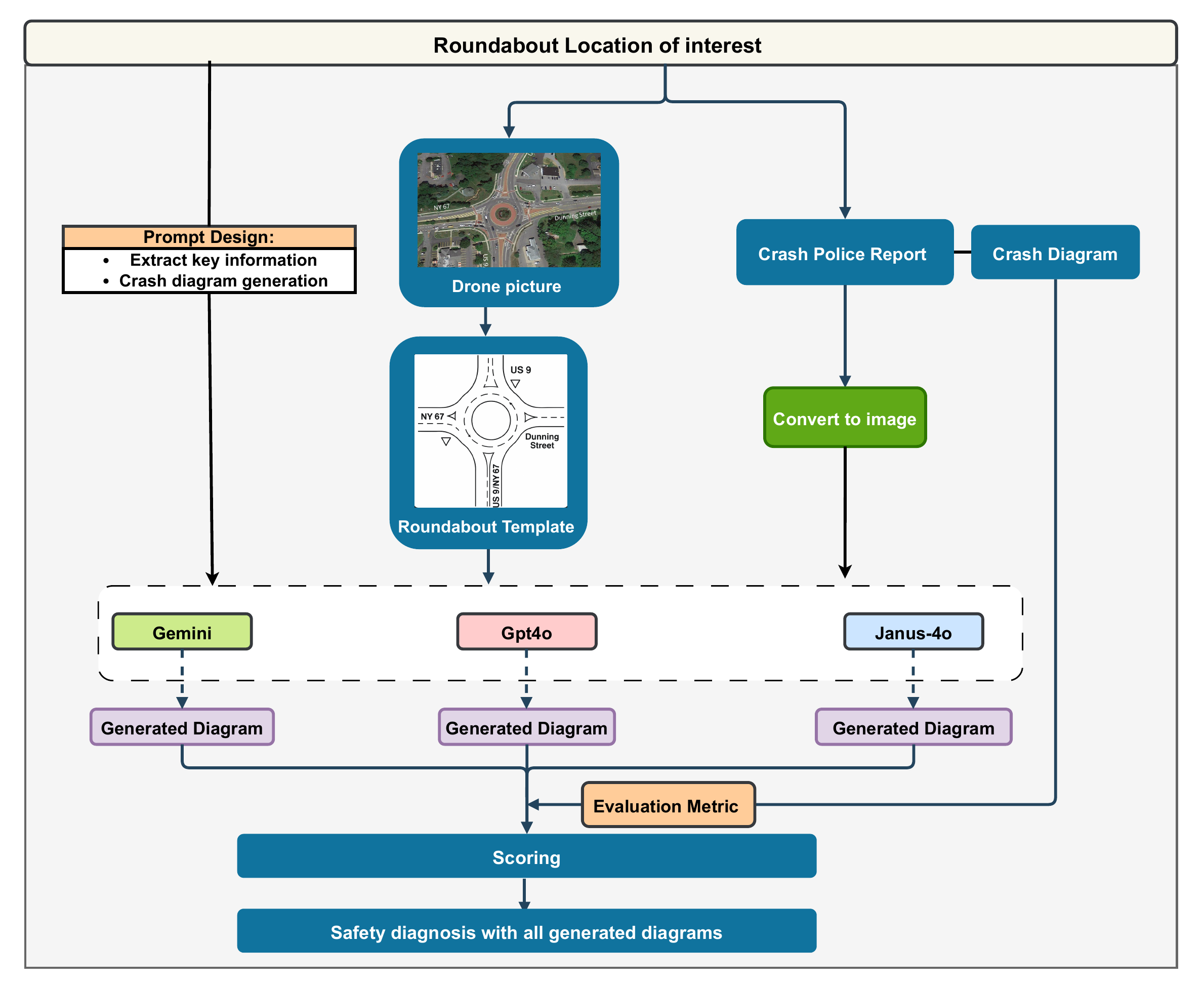}
    \caption{Crash diagram generation workflow using drone-based aerial imagery and vision-language models.}
    \label{fig:workflow}
\end{figure}

The proposed crash diagram generation workflow, as delineated in Figure~\ref{fig:workflow}, establishes a comprehensive methodological framework that synthesizes heterogeneous data modalities through advanced vision-language model architectures to achieve automated crash visualization synthesis. The methodology commences with the strategic identification of roundabout intersections as focal points of analysis, whereupon high-resolution aerial imagery collected by drones is systematically processed into standardized geometric templates to ensure precise spatial referencing of the roundabout base image. 

The central component of this framework involves a structured prompt engineering approach that systematically extracts key incident information directly from digitized crash report images while providing clear instructions for diagram generation. This organized prompt structure, working together with the geometric template system, serves as input to three leading vision-language models, Gemini-1.5-Flash~\cite{anil2023gemini}, GPT-4o\cite{openai2024chatgpt4o}, and Janus-4o\cite{chen2025janus4o}, each operating independently to generate crash diagrams based on the provided information.

The framework includes a specialized evaluation system designed specifically for assessing crash diagram quality, enabling systematic comparison of generated outputs against ground truth diagrams created by qualified traffic safety experts. This complete computational pipeline allows for thorough comparison of model performance while maintaining consistent standards across data preprocessing, prompt design, and evaluation methods, thereby contributing to both theoretical understanding of multimodal generation capabilities and practical applications in automated safety visualization systems.

\subsection*{Problem formulation}

The objective of this study is to develop a vision-language framework that generates spatially accurate crash diagrams based on structured prompts and visual references. The task fundamentally is a images-text to image problem with vision-language models. Each input consists of a detailed instructional prompt describing the crash diagram task, along with a standardized roundabout layout image and targeted digitized crash report image. The desired output is a diagram that visually reconstructs the crash scenario with correct vehicle positioning, movement, and point of impact.

Let 
\begin{itemize}
    \item $P$ denote the structured instructional prompt, containing narrative, road labels, vehicle behaviors, crash type definitions, and damage code interpretation rules.
    \item $I_r$ represent the reference image of a multilane roundabout (used consistently across all cases).
    \item $I_c$ represent the targeted digitized crash report image containing specific incident details.
    \item $D$ denote the generated crash diagram, which must depict vehicle paths, impact points, and an information box.
\end{itemize}

The problem can be formally defined as a multimodal mapping function:
\[
\boldsymbol{f: (P, I_r, I_c) \longrightarrow D},
\]
where $f$ is implemented by a vision-language model capable of aligning visual spatial reasoning with structured textual instructions and crash report information.
The prompt input $P$ can be decomposed into the following components:
\[
\boldsymbol{P = \{T, C, G\}},
\]
where 
\begin{itemize}
    \item $T$ is the crash narrative and intersection location,
    \item $C$ is collision metadata, including Box 1 damage codes for each vehicle,
    \item $G$ is a graphical task description defining diagram requirements and layout constraints.
\end{itemize}

The output diagram $D$ must satisfy several critical spatial and informational accuracy requirements to ensure valid crash reconstruction. Vehicle positioning must adhere to the geometric constraints of the roundabout infrastructure, with each vehicle placed within the appropriate circulating lanes corresponding to their documented movements and designated approach roads. The spatial representation of the collision point must demonstrate precise alignment with the reported damage codes extracted from the crash report, ensuring that the impact location reflects the physical evidence documented in the incident.

Furthermore, the diagram must provide clear demarcation of vehicle entry and exit trajectories, with accurate vehicle identification labels (V1, V2) that correspond to the entities described in the crash report. The visualization framework requires the inclusion of a comprehensive information summary box that synthesizes the collision classification, vehicle damage coding, and a concise narrative description of the incident sequence, thereby providing both visual and textual documentation of the crash scenario within a unified representational framework.

This formalization establishes the task as a structured vision-language generation problem that transforms multimodal inputs into consistent and interpretable crash diagrams suitable for traffic safety analysis. The present study employs roundabout crash diagram generation as the primary evaluation domain to assess the effectiveness of this methodological approach. Previous research in automated crash analysis has largely applied natural-language processing to text fields or computer-vision algorithms to detect events in images and video. However, few studies have investigated the generation of interpretable engineering diagrams directly from textual police reports. Recent advances in multimodal learning and structured scene synthesis (6–8) suggest that VLMs possess the cross-modal reasoning capacity needed for such tasks, motivating this study’s focus on evaluating their feasibility in traffic safety documentation.

\subsection*{Vision-language models for image generation}

This study investigates the utilization of Vision-Language Models (VLMs) to automate the generation of traffic crash diagrams from structured police report data. Specifically, three advanced multimodal models, including OpenAI’s GPT-4o, Google DeepMind's Gemini-1.5-Flash, and the open-source Janus-4o, were assessed for their capability to extract essential information from crash report, understand the instructions, and produce spatially accurate crash diagrams within multilane roundabouts.

\textbf{GPT-4o} (short for GPT-4 Omni) constitutes OpenAI’s latest advancement in unified multimodal processing. This model is based on a transformer architecture that simultaneously encodes and generates content across both text and image modalities. GPT-4o is specifically optimized for high-fidelity visual tasks requiring detailed spatial reasoning, including diagram synthesis, visual question answering, and scene understanding. Its capacity to interpret structured prompts and produce semantically coherent, top-down visuals makes it particularly suitable for structured crash documentation tasks, where strict adherence to spatial logic and lane discipline is critical.

\textbf{Gemini} is Google DeepMind's premier multimodal model family, engineered to perform complex reasoning across a variety of input types, including text, images, audio, and video. Gemini utilize a closely integrated vision encoder and language decoder architecture, facilitating advanced cross-modal reasoning. For this task, Gemini’s strengths lie in interpreting richly structured prompts and converting textual descriptions into visually coherent outputs. Its proficiency in visual rendering, especially in producing clean diagram layouts and maintaining consistent labeling, establishes a robust baseline for comparative assessments.

\textbf{Janus-4o} is an emerging open-source multimodal large language model (MLLM) derived from Janus-Pro-7B and fine-tuned on the ShareGPT-4o-Image dataset. In contrast to previous versions of Janus, Janus-4o supports both text-to-image and text-and-image-to-image generation, demonstrating significant enhancements in visual outputs. Although it remains under active development, Janus-4o seeks to approximate the image synthesis capabilities of GPT-4o through parameter-efficient fine-tuning techniques. In this study, Janus{-4o} was evaluated to determine the extent to which an open-source MLLM can perform comparably to proprietary state-of-the-art systems.

{The experiments were conducted using the following model versions: GPT-4o (API version: gpt-4o-2024-05-13), Gemini-1.5-Flash, and Janus-4o. Default API parameters were used unless otherwise specified.}
The task for all three models involved converting structured crash report inputs into spatially accurate crash diagrams that visually represent vehicle interactions within a two-lane roundabout environment. 
Each input consisted of three components: a detailed prompt derived from the police crash report, which included narrative descriptions, directional information, damage codes, vehicle paths, and collision classifications; a reference base layout of the roundabout, including consistent geometric features such as circulating lanes, entry and exit points, and standardized road orientations (e.g., North, East, South, West); and the digitized crash report image containing the original incident documentation and specific crash details for multimodal processing.

The intended output was a top-down crash diagram that accurately reconstructed the incident with spatial fidelity. Specifically, the diagrams were expected to 

\begin{itemize}
\item Accurately place each vehicle within the circulating lanes (not the center island);
\item Reflect correct entry and exit paths according to the described movements;
\item Indicate the precise point of impact based on reported damage codes;
\item Label vehicles clearly as V1 and V2;
\item Include a narrative box summarizing the crash type, damage codes, and sequence of events.
\end{itemize}

Unlike open-ended text-to-image generation tasks, this scenario required constrained generation limited by roadway geometry, traffic control conventions, and crash reconstruction principles. Consequently, it functioned as a rigorous benchmark for assessing each model's ability to translate multimodal, semantically rich, and spatially explicit crash report content into diagrammatic form. Ultimately, this task underscores the potential role of VLMs in automating elements of crash analysis that have traditionally handled by trained human analysts.

\section{Data and experiments}
This section describes the crash dataset, structured prompt design, model inference process,
and evaluation methodology used to assess vision-language model performance.
{This section describes the crash dataset, structured prompt design, model inference process,
and evaluation methodology used to assess vision-language model performance.}

\subsection{Roundabout data description}

The crash dataset analyzed in this study was obtained from official New York State Police Accident Reports (Form MV-104A), as part of the ``Reasons for Drivers Failing To Yield at Multi-Lane Roundabout Exits: Transportation Pooled Fund Study''~\cite{Medina2023}. This research specifically focused on crashes occurring at two high-volume multilane roundabouts in the Town of Malta, New York. These roundabouts are located at the intersections of U.S. Route 9 and New York State Route 67, and Route~67 and Dunning Street. These sites are critical nodes within the local transportation network, located in a suburban commercial context that accommodates both commuter and freight traffic.

\begin{figure}[!htbp]
	\centering
	\includegraphics[width=0.5\textwidth]{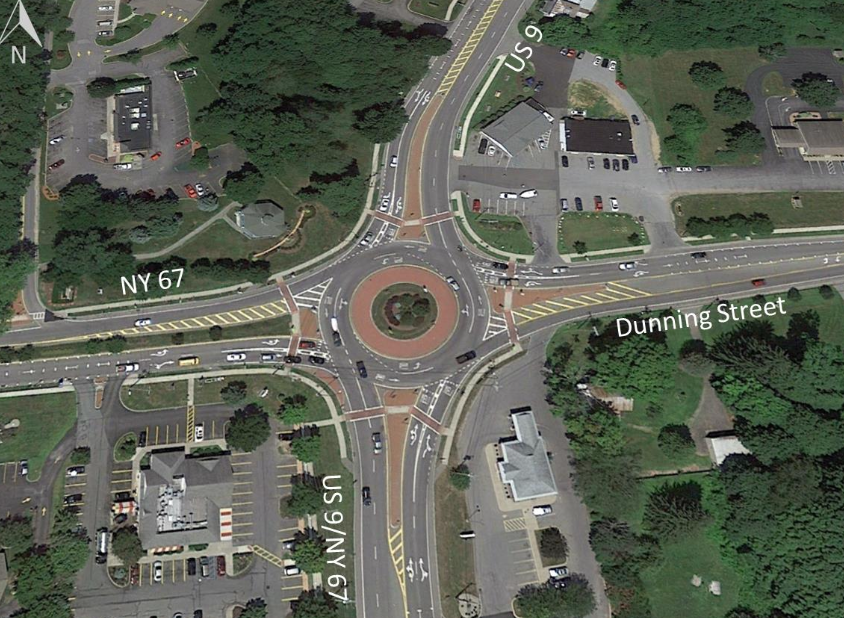}
	\caption{Aerial view of the study roundabout.}
	\label{fig:studyroundaboutdiagram}
\end{figure}

The study roundabout show in Figure \ref{fig:studyroundaboutdiagram} experiences an average daily traffic volume of approximately 26,500 vehicles. The roundabout geometry features a two-lane circulating roadway, with each approach directions: northbound (US 9), southbound (US 9/US 7), eastbound (Dunning Street), and westbound (NY 67). The roundabout features an inscribed circle diameter of 165 feet and a central island diameter of 105 feet. The circulatory roadway maintains a consistent width of 30 feet, divided into two lanes each measuring 15 feet in width. Entry and exit lanes vary between 28 and 31 feet in width, with individual entry lanes typically ranging from 12 to 13 feet.

The presence of spiral pavement markings, which serve to guide vehicles, particularly those making left turns, from the inside circulating lane to the correct exit point. These spirals are essential due to the single-lane exit configurations on both the eastbound (Dunning Street) and westbound (NY~67) legs. However, the markings were introduced retrospectively, the resulting spiral geometry is relatively abrupt. Field observations indicate that such conditions contribute to driver confusion and pose potential safety risks associated with improper lane changes and last-minute maneuvers. While the study sites each contain four major approaches, the proposed framework can be generalized to roundabouts with a greater number of entries or exits by adjusting the base geometric template and prompt instructions accordingly.

A total of 79 crash reports were selected for inclusion in the dataset. These reports were chosen based on the completeness of key elements required for subsequent analysis and diagram generation. The inclusion criteria required the presence of:  a detailed narrative description of the collision, directional indicators for each involved vehicle, clearly identified vehicle damage codes located in ``point of Impact'', and accompanying officer-drawn crash sketches. The selected reports represent a range of crash types, including rear-end collisions, sideswipes, improper merging, overtaking, and failure-to-yield incidents—many of which are known to be exacerbated by complex lane configurations at multilane roundabouts.

Each MV-104A report contains structured fields capturing vital crash descriptors, including:
\begin{itemize}
\item Vehicle trajectories and movement patterns;
\item Collision narratives and driver violation codes;
\item Time of day, weather, lighting, and roadway surface conditions;
\item Impact locations coded using a standardized numeric system.
\end{itemize}

\begin{figure}[!htbp]
	\centering
	\includegraphics[width=0.7\textwidth]{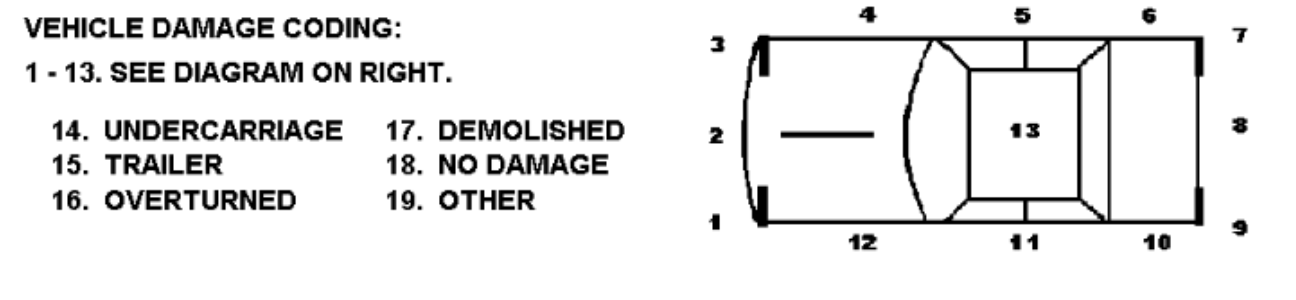}
	\caption{New York state DMV damage code diagram for vehicle impact regions.}
	\label{fig:damage-diagram}
\end{figure}

A fundamental aspect of this study entailed decoding the numeric vehicle damage codes documented in “Box 1–Point of Impact”. These codes specify the precise areas of a vehicle that incurred damage during a collision and are critical for verifying the accuracy of the generated crash diagrams. Codes 1 through 13, as described in Table 1, correspond to fixed locations on the vehicle body, as illustrated in Figure~\ref{fig:damage-diagram}. Additional codes ranging from 14 to 19 account for less localized or more generalized outcomes such as undercarriage damage, overturned vehicles, total demolition, and instances where no visible damage was reported.

\begin{table}[h!]
\centering
\caption{Vehicle damage codes and descriptions.}
\label{tab:damage-codes}
\begin{tabular}{ll}
\hline
\textbf{Code} & \textbf{Description} \\
\hline
1  & Left Front Corner        \\
2  & {Front Center}             \\
3  & {Right Front Corner}       \\
4  & {Right Front Fender}           \\
5  & {Right Side Door}         \\
6  & {Right Rear Fender}              \\
7  & {Right Rear Light}        \\
8  & {Rear Trunk Center}          \\
9  & {Left Rear Light}        \\
10 & {Left Rear Fender}         \\
11 & {Left Side Door}       \\
12 & {Left Front Fender}             \\
13 & Top or Undercarriage     \\
\hline
\end{tabular}
\end{table}

\subsection{Prompt construction and evaluation metric}

To produce precise and standardized crash diagrams from police crash reports, a structured prompt was designed and implemented for use with vision-language models (VLMs). This prompt was manually developed and aligned with a consistent template to ensure the reliable transfer of essential contextual, spatial, and semantic information from the MV-104A form to the diagramming model. The prompt was divided into three logically sequenced sections to systematically guide the model through scenario interpretation, information extraction, and final output synthesis.

The first section of the prompt presented the fixed geometric configuration of the study roundabout and outlined the fundamental instructions for diagram construction. This section stated:

\begin{quote}
“You are provided with a traffic crash report and a standard roundabout layout. Your task is to extract key information and generate a standardized crash diagram that accurately represents the crash scenario on the roundabout.

\textbf{Roundabout Layout (Always Use This Configuration):}  
Northbound: US 9  
Eastbound: Dunning Street  
Southbound: US 9 / US 7  
Westbound: NY 67  
The roundabout has two circulating lanes.  
Vehicles must be shown on appropriate lanes within the roundabout (not the center).  

\textbf{Crash Diagram Requirements:}  
Place and label vehicles involved in the crash as V1 and V2.  
Show direction of travel and point of impact.  
Accurately depict entry and exit points for each vehicle.  
Include a clearly labeled information box in the bottom-left corner, showing:  
Collision Type (one of the nine standard types)  
V1 Damage Code (from `Box 1–Point of Impact')  
V2 Damage Code (from `Box 1–Point of Impact')  
Narrative Summary (a concise sequence of events describing the crash)”
\end{quote}

This section of the prompt was essential in establishing a consistent visual framework for the model to adhere to. By specifying fixed compass orientations and the geometric layout of the roundabout, spatial consistency was maintained across all generated diagrams. Explicit instructions to position vehicles within appropriate lanes of the roundabout (excluding the center) minimized unrealistic placements. Additionally, directives to label vehicles as “V1” and “V2”, indicate movement directions, and mark impact points facilitated meaningful diagrammatic interpretation of the crash scenario. The inclusion of a bottom-left information box further reinforced the semantic structure by linking the diagrammatic visuals to structured crash data.

The second part of the prompt directed the model to extract relevant data from the police report and apply it within the context of the roundabout. It stated:
\begin{quote}
``From the Crash Report, Extract and Interpret:  
\textbf{Location/Intersection Name}  
Confirm the crash occurred at the roundabout.  
Identify all approach roads and confirm their orientations with the layout directions.  

\textbf{Collision Type}  
Classify the crash as one of the following types:  
Rear-End, Overtaking, Right Turn, Left Turn, Right Angle, Head-On, Right Turn (variation), Sideswipe  

\textbf{Vehicle Movement and Positioning}  
Determine where each vehicle entered and exited the roundabout.  
Show the correct movement path and trajectory within the roundabout.  
Ensure proper lane usage consistent with a two-lane roundabout.  

\textbf{Narrative Summary}  
Convert the crash narrative into a clear and concise chronological sequence.  
Example: ‘V1 entered the roundabout from Dunning Street (eastbound) and failed to yield to V2 already circulating from US 9/US 7 (southbound)’.  

\textbf{Damage Codes}  
Extract numeric damage codes (1–13) from ‘Box 1–Point of Impact’ for both vehicles.  
Use the following guide to interpret the codes:  
1 = Left Front Corner (Driver's headlight)  
2 = Front Center (front bumper center)  
3 = Right Front Corner (Passenger headlight)  
4 = Right Front Fender  
5 = Right Side Door  
6 = Right Rear Fender  
7 = Right Rear Light  
8 = Rear Trunk Center  
9 = Left Rear Light  
10 = Left Rear Fender  
11 = Left Side Door  
12 = Left Front Fender  
13 = Roof / Hood / Trunk Top”
\end{quote}

This section operationalizes the interpretation of structured police report fields, including spatial positioning, coded impact zones, and narrative events. It explicitly directs the model to verify crash location and directionality by cross-referencing approach road names with compass-based orientation. The classification of collision types into discrete categories ensures terminological consistency, while instructions for movement interpretation reinforce lane-specific vehicle trajectories. Additionally, narrative conversion and damage code interpretation facilitate the translation of semi-structured textual inputs into precise visual representations, grounded in the official DMV reference system.

The third and final section of the prompt delineated the expectations for the synthesized crash diagram, emphasizing the importance of completeness, clarity, and diagrammatic integrity. This section stated:
\begin{quote}
“Final Output:  
A visually accurate roundabout crash diagram that shows:  
Proper road geometry and vehicle placements  
Entry/exit points and travel directions  
Points of impact  
An information box summarizing extracted crash data”
\end{quote}

This concluding directive integrated all previous prompt elements and restated the essential output requirements. It emphasized that the diagram should demonstrate both spatial accuracy,  such as precise road geometry and trajectory positioning, and semantic clarity by incorporating an information summary box. These criteria are vital to ensure consistent evaluation across multiple models and reports.

Collectively, these three sections of the prompt constituted a comprehensive set of instructions for vision-language models to interpret complex crash reports and generate accurate, interpretable visual representations. This structured prompt design ensured alignment between the report content and the produced diagrams, thereby facilitating robust downstream evaluation across multiple models and metrics.

Following the development of structured prompts, each prompt was presented to three VLMs (GPT-4o, Gemini-1.5-Flash, and Janus-4o) through a scripted API pipeline designed to maintain consistent input formatting and enable automated batch processing. Each model produced a schematic crash diagram featuring standardized top-down geometry, directional arrows, labeled vehicles (V1 and V2), and a narrative information box summarizing the crash context and damage details. This procedure was applied to all 79 crash reports, yielding a uniform dataset of model-generated diagrams for comparative analysis.

\subsection{Evaluation metrics design}
To systematically assess the effectiveness of each model in translating structured crash data into visually accurate representations, we developed a comprehensive evaluation metric set based on established traffic safety standards for crash diagram, as shown in Table \ref{tab:metric-description}. This metric set was designed to quantify both the semantic accuracy and spatial fidelity of the generated diagrams relative to the original MV-104A crash diagrams in the crash reports.

\begin{table}[!htbp]
	\centering
	\caption{Crash diagram evaluation metrics.}
	\label{tab:metric-description}
	\begin{tabular}{lp{9.7cm}}
		\hline
		\textbf{Metric} & \textbf{Description} \\
		\hline
		Collision Type Extraction & Correct collision type (e.g., rear-end, angle) is accurately extracted from the report and depicted in the diagram. \\ 
		
		Vehicle Labeling–V1 & Vehicle 1 is clearly labeled as ``V1" in the crash diagram. \\ 
		
		Vehicle Labeling–V2 & Vehicle 2 is clearly labeled as ``V2" in the crash diagram. \\ 
		
		Collision Location & The crash is depicted in the correct quadrant or at the correct entry/exit point of the roundabout. \\ 
		
		Collision Point Accuracy & The exact spot of vehicle impact is accurately marked in the diagram. \\ 
		
		V1 Damage Code–Extraction Accuracy & The damage code for Vehicle 1 is accurately extracted from the crash report. \\ 
		
		V1 Damage Code–Visual Consistency & The damage on Vehicle 1 is visually consistent with the extracted damage code (e.g., left front corner). \\ 
		
		V2 Damage Code–Extraction Accuracy & The damage code for Vehicle 2 is accurately extracted from the crash report. \\ 
		
		V2 Damage Code–Visual Consistency & The damage on Vehicle 2 is visually consistent with the extracted damage code (e.g., right rear side). \\ 
		
		Overall Clarity and Proportion & The roundabout layout, lane markings, and vehicle illustrations are proportionate and clearly presented. \\ \hline
	\end{tabular}
\end{table}


A strict binary (0/1) scoring scheme was adopted because crash diagrams are safety-critical engineering artifacts, where partial correctness may still lead to misinterpretation.
This approach emphasizes exact spatial and semantic accuracy while enabling consistent
comparison across models and evaluators. 
The scoring rubric was developed, comprising ten specific evaluation metrics. These metrics were selected to assess critical aspects of diagram quality, including the accuracy of collision type, vehicle labeling, trajectory representation, impact location, and alignment with coded damage information. Each metric was scored as 1 for accurate representation or 0 for missing, incorrect, or ambiguous elements, yielding a maximum possible score of 10 for each diagram.

{All diagrams were independently evaluated by trained human experts with experience in crash analysis and diagram interpretation. Any scoring disagreements were resolved through discussion to ensure consistent application of the evaluation rubric.} 
Each evaluator independently reviewed every diagram alongside its corresponding crash report and constructed prompt to assess the alignment between the textual inputs and visual outputs. This rigorous comparative methodology enabled a comprehensive analysis of model performance, highlighting strengths and limitations in spatial layout, vehicle dynamics, and visual clarity. The structured rubric combined with an expert-led evaluation process facilitated consistent scoring across various models and crash types, thereby establishing a foundation for performance benchmarking and comparative error analysis.

The inclusion of standard deviations across the ten metrics indicated stable model behavior, with most values within $\pm 0.10$ for GPT-4o and $\pm 0.18$ for Gemini-1.5-Flash, while Janus-4o exhibited higher variability. GPT-4o’s consistency across semantic and spatial metrics suggests a stronger alignment between textual comprehension and diagram generation. Gemini-1.5-Flash showed moderate but uneven performance across categories, whereas Janus-4o’s output variability reflected difficulties in multimodal reasoning and geometric precision.

\section{Results and discussion}

This section presents a comparative analysis of the performance of three popular VLMs (GPT-4o, Gemini-1.5-Flash, and Janus-4o) in generating roundabout crash diagrams from standard police report. Each model was evaluated using a ten-point binary metric system designed to assess key aspects of diagram accuracy, including semantic extraction (e.g., collision type), spatial reasoning (e.g., quadrant and point of impact), damage localization, and overall visual clarity. The analysis is based on a dataset comprising 79 unique crash reports, with each model producing a corresponding diagram for each report. 

Table~\ref{tab:metric-avg} presents the average scores obtained by each model for various evaluation metrics across all 79 diagrams. Metric values exceeding 0.90 are highlighted in green to indicate high consistency, whereas scores below 0.30 are shaded in red to signify areas of weakness. GPT-4o outperformed both Gemini-1.5-Flash and Janus{-4o} on nearly all metrics, particularly in critical domains such as collision type identification, labeling consistency, and clarity. Gemini-1.5-Flash exhibited moderate performance, with notable deficiencies in spatial accuracy and damage localization. Janus-4o showed the weakest overall performance, especially in the interpretation of semantic elements and the localization of damage codes. {Model performance is summarized using per-metric average scores across the evaluated crash cases under a strict binary (0/1) scoring scheme. Performance variability is characterized using per-metric standard deviations computed across crash cases for each model, where binary scores (0/1) assigned by expert reviewers are used to calculate the standard deviation for each evaluation metric. The resulting standard deviations indicate relatively stable behavior for GPT-4o, with most values within approximately $\pm$0.10, and moderate variability for Gemini-1.5-Flash, with most values within approximately $\pm$0.18, while Janus-4o exhibits higher variability across metrics. While variability across crash scenarios was observed, formal confidence intervals for aggregate scores were not computed in this initial benchmark, as the primary goal is to establish a first evaluation framework for this previously unexplored task.}

\begin{table}[!htbp]
\centering
\caption{Average metric scores by model.}
\label{tab:metric-avg}
\begin{tabular}{llll}
\hline
\textbf{Metric} & \textbf{GPT-4o} & \textbf{Gemini-1.5-Flash} & \textbf{Janus{-4o}} \\
\hline
Collision Type Extraction & \cellcolor{green!20}0.95 & 0.70 & \cellcolor{red!20}0.33 \\
Vehicle Labeling–V1     & \cellcolor{green!20}0.98 & \cellcolor{green!20}0.93 & \cellcolor{green!20}1.00 \\
Vehicle Labeling–V2     & \cellcolor{green!20}0.91 & \cellcolor{green!20}0.98 & \cellcolor{green!20}1.00 \\
Collision Location        & 0.42 & 0.44 & \cellcolor{red!20}0.09 \\
Collision Point Accuracy  & \cellcolor{red!20}0.16 & \cellcolor{red!20}0.02 & \cellcolor{red!20}0.00 \\
V1 Damage Code–Extraction Accuracy & 0.47 & \cellcolor{red!20}0.19 & \cellcolor{red!20}0.07 \\
V1 Damage Code–Visual Consistency  & \cellcolor{red!20}0.30 & \cellcolor{red!20}0.26 & \cellcolor{red!20}0.00 \\
V2 Damage Code–Extraction Accuracy & 0.44 & \cellcolor{red!20}0.07 & \cellcolor{red!20}0.00 \\
V2 Damage Code–Visual Consistency  & 0.42 & \cellcolor{red!20}0.16 & \cellcolor{red!20}0.00 \\
Overall Clarity and Proportion       & \cellcolor{green!20}0.91 & \cellcolor{green!20}0.93 & \cellcolor{green!20}0.98 \\
\hline
\textbf{Total Score (out of 10)}     & \textbf{6.29} & \textbf{5.28} & \textbf{3.64} \\
\hline
\end{tabular}
\end{table}

While GPT-4o and Gemini-1.5-Flash demonstrated comparable performance in vehicle labeling and diagram clarity, GPT-4o exhibited a marked advantage in tasks necessitating reasoning across both textual and spatial domains. Notably, its consistent ability to extract damage codes and accurately align them with visual representations indicates a more robust understanding. In contrast, Janus-4o, despite maintaining visual consistency, failed to capture essential elements of the crash narrative, instead relying on template-based outputs with limited contextual adaptation. {Despite GPT-4o achieving the highest overall performance, persistent weaknesses were observed in collision point accuracy and damage-code visual consistency. Misplacement of impact points and incomplete damage visualization remain common failure modes, indicating that current vision-language models do not yet achieve the spatial precision required for fully autonomous crash documentation. As a result, the proposed framework should be viewed as a decision-support tool that can enhance efficiency and consistency, while continued human oversight remains essential, particularly for applications with operational or legal implications.}

{Low scores for collision location and collision point accuracy indicate that current vision-language models struggle with precise spatial reasoning and geometric fidelity, which are essential for engineering-grade diagram generation. While the proposed framework demonstrates the feasibility of automated crash diagram synthesis, the results indicate that current vision-language models remain limited in precise spatial coordinate mapping and geometric fidelity. As such, the approach is not yet suitable for fully automated engineering-grade applications.}

A visual examination of the generated diagrams further supports the observed quantitative trends. The diagrams produced by GPT-4o consistently depicted accurate collision dynamics, including precise lane positioning and direction of travel. In contrast, the outputs from Gemini-1.5-Flash, while often visually clean, lacked substantive detail. Gemini-1.5-Flash frequently omitted damage placement and encountered difficulties in accurately extracting damage codes, which could result in misinterpretations of vehicle positioning. Janus-4o exhibited limited capability in extracting both damage codes and their locations. Although it correctly labeled vehicles and maintained consistent formatting, its diagrams failed to accurately represent the reported crash information.

\begin{figure}[!htbp]
	\centering
	\begin{subfigure}[b]{0.45\textwidth}
		\centering		
		\includegraphics[width=0.9\textwidth]{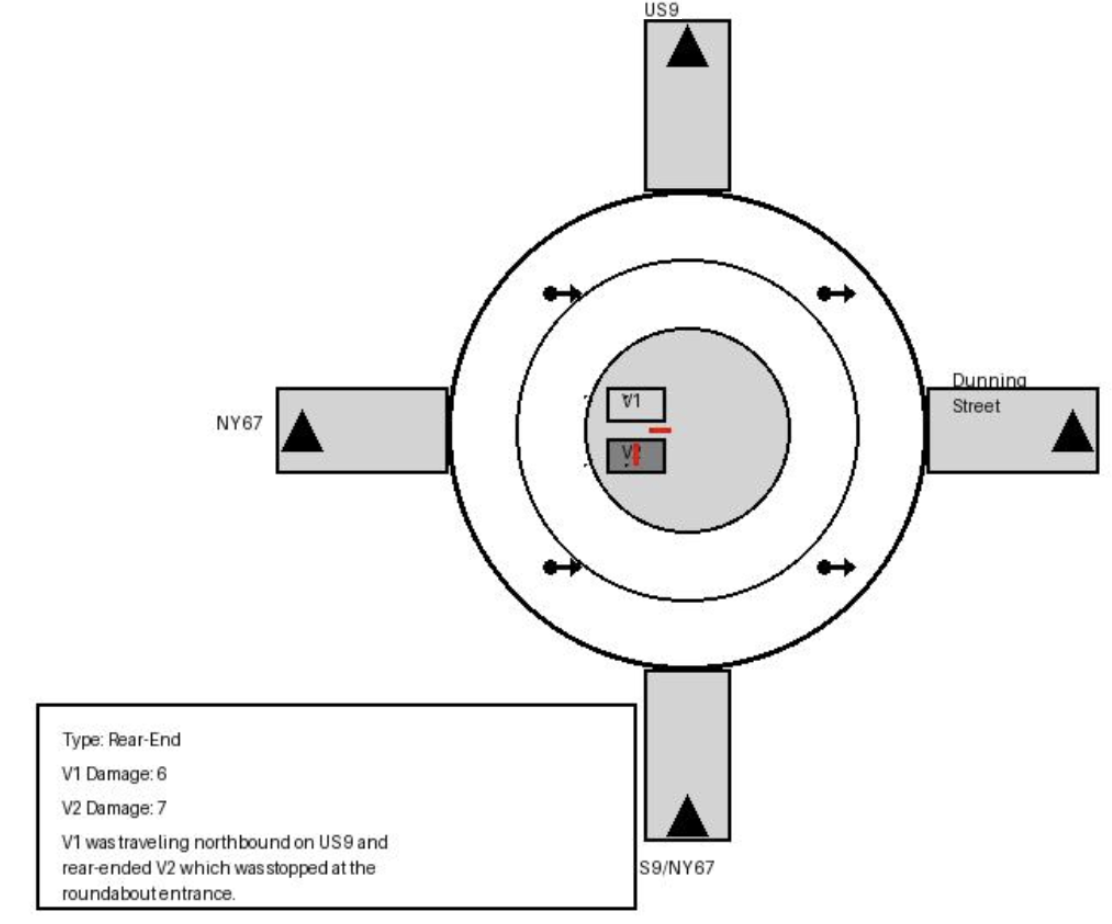}\vspace{-2mm}
		\subcaption{Generated crash diagram by Janus-4o}
		\label{fig:janus1}
	\end{subfigure}\hspace{0.4in}
	\begin{subfigure}[b]{0.45\textwidth}
		\centering
		\includegraphics[width=0.9\textwidth]{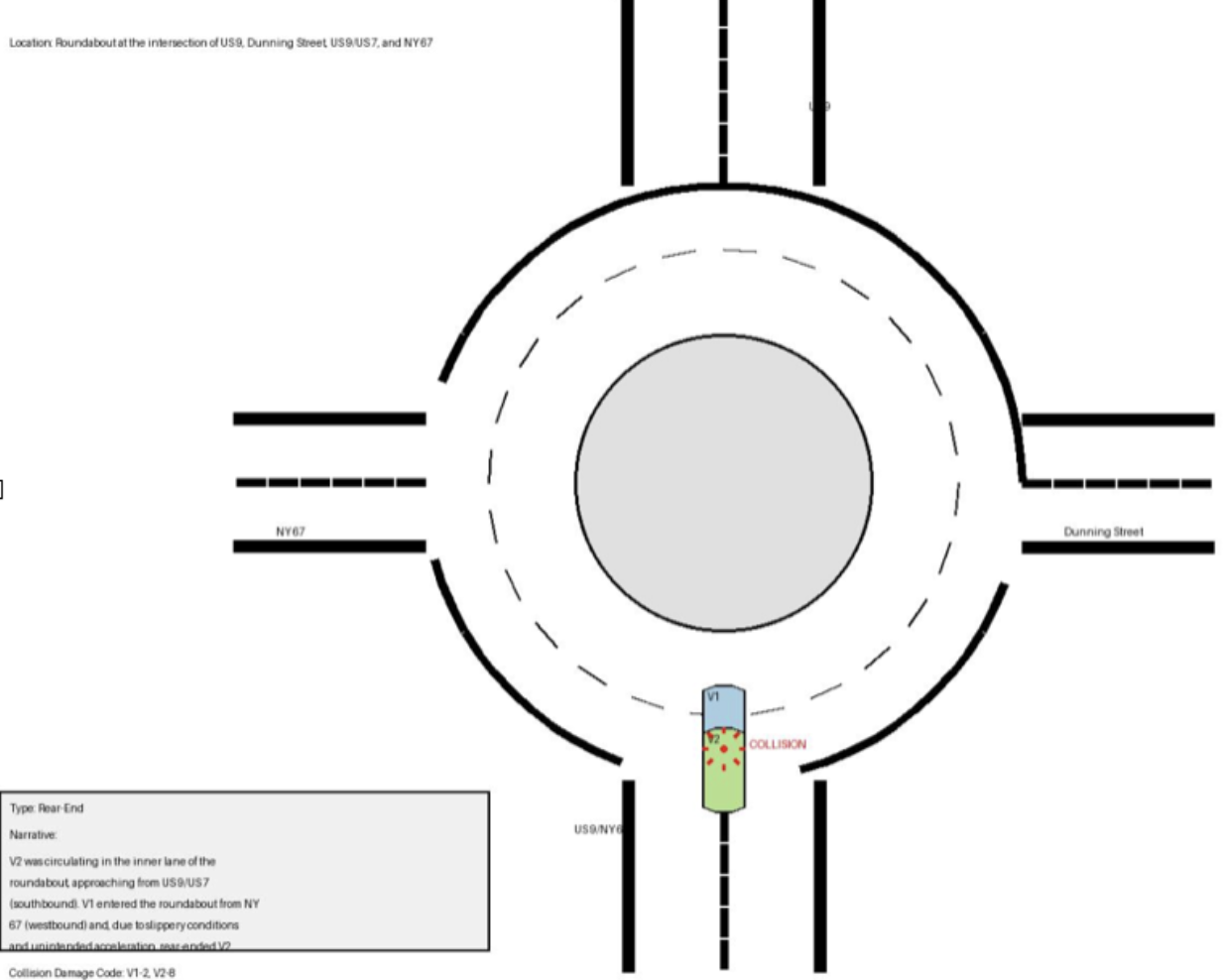}\vspace{-2mm}\subcaption{Generated crash diagram by Gemini-1.5-Flash}
		\label{fig:gemini1}
	\end{subfigure}
	\begin{subfigure}[b]{0.45\textwidth}
		\centering
		\includegraphics[width=0.9\textwidth]{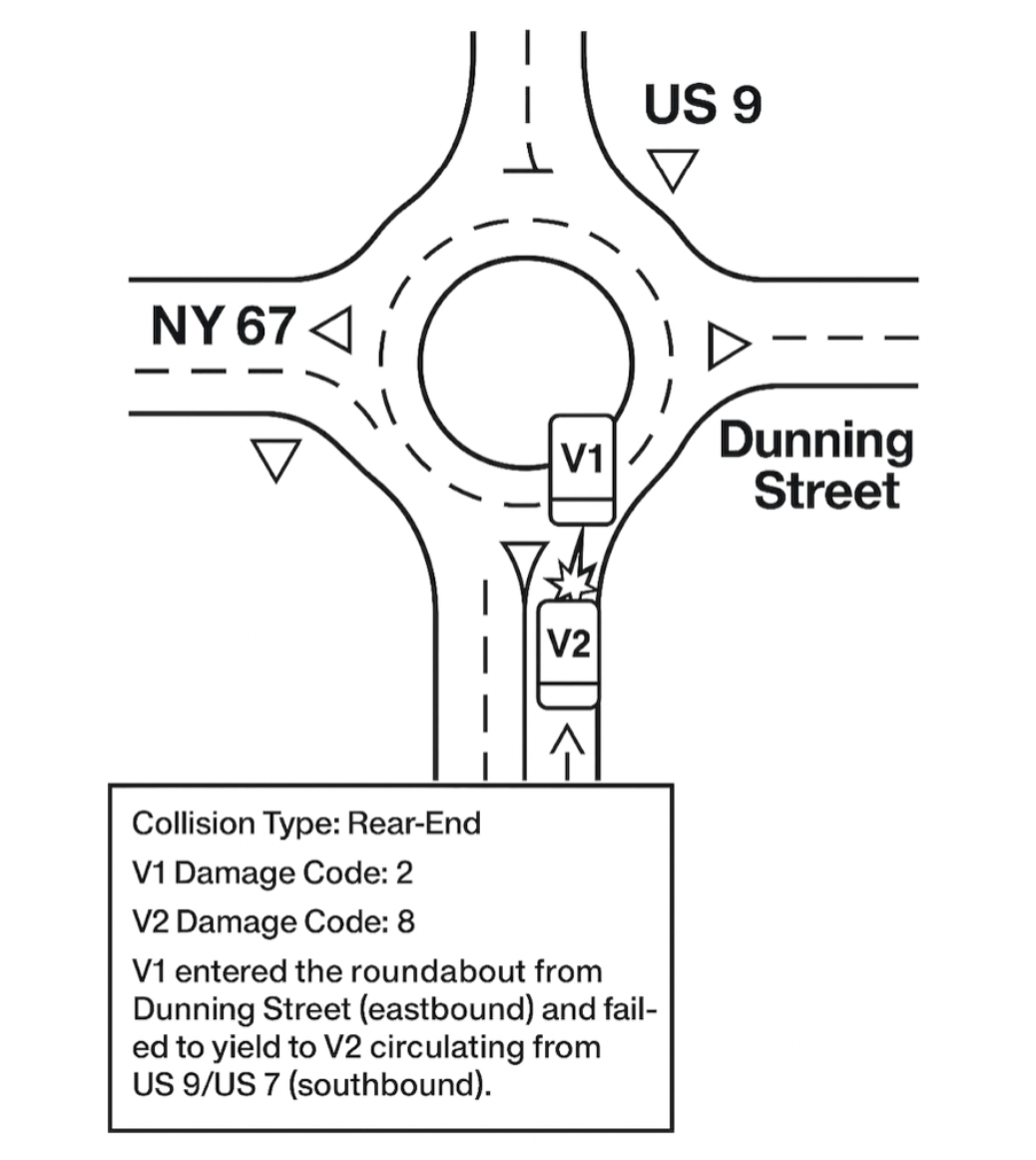}\vspace{-2mm}
		\subcaption{Generated crash diagram by GPT-4o}
		\label{fig:chatgpt1}
	\end{subfigure}\hspace{0.4in}
	\begin{subfigure}[b]{0.45\textwidth}
		\centering
		\includegraphics[width=0.9\textwidth]{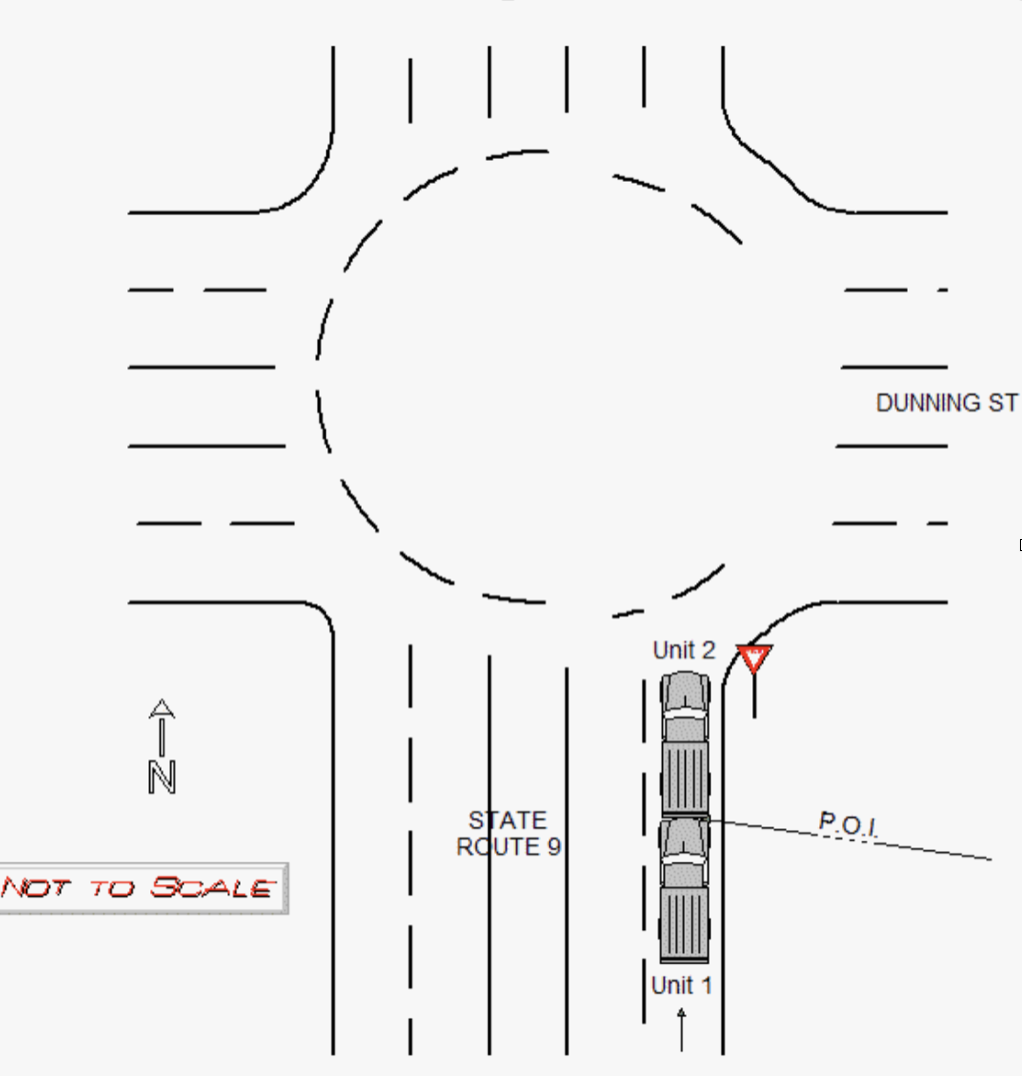}\vspace{-2mm}
		\subcaption{Crash diagram - ground truth (police report)}
        		\label{fig:police1}
	\end{subfigure}	\vspace{-2mm}
	\caption{Comparison of crash diagrams generated by Janus-4o, Gemini-1.5-Flash, and GPT-4o against the official police crash diagram for the same crash.}
	\label{fig:comparison}
\end{figure}

Figure \ref{fig:comparison} provides a comparative analysis of crash diagrams produced by three Vision-Language Models: Janus-4o (top-left), Gemini-1.5-Flash (top-right), and GPT-4o (bottom-left)—alongside the official police crash diagram (bottom-right), all generated from the same structured crash report. This comparison elucidates the respective strengths and limitations of each model in terms of spatial reasoning, semantic extraction, and adherence to templates.

The diagram generated by Janus{-4o} (top-left) does not successfully reconstruct a coherent roundabout configuration. It lacks identifiable circular geometry and a clear layout of road connections. Furthermore, both vehicles are positioned within the central island instead of the circulating lanes. Essential elements, including the accurate quadrant location, collision type, and damage code, are either absent or inaccurately represented.

The output produced by Gemini-1.5-Flash (top-right) exhibits certain improvements. It accurately positions a vehicle within the circulating lane and correctly identifies the collision in the appropriate quadrant. However, only one vehicle is distinctly visible, while the existence of a second vehicle is suggested but not explicitly illustrated. The spatial relationship between the vehicles remains ambiguous due to partial occlusion, and the impact zone lacks a clearly defined collision marker. Although the diagram maintains closer conformity to the base roundabout layout compared to Janus-4o, it is insufficient in effectively conveying the interaction between the two vehicles involved in the crash.

In contrast, GPT-4o (bottom-left) demonstrates enhanced spatial reasoning and greater fidelity to both the roundabout template and the narrative content provided in the input prompt. The roundabout is accurately depicted with two circulating lanes and clearly indicated cardinal directions. Vehicles V1 and V2 are correctly positioned within the circulating lanes, with directional arrows denoting their approach from Dunning Street and US 9, respectively. The diagram also presents a plausible crash configuration, featuring an accurate point of impact and appropriately assigned damage codes. Furthermore, the narrative summary and vehicle labels are legible and consistent with the report content, rendering this output the most congruent with the expected diagram format.

The official police diagram (located at the bottom right) functions as the evaluation baseline. It features a comprehensive and proportionally accurate roundabout layout, annotated vehicle trajectories, directional arrows, labeled impact points, and clearly indicated damage codes. This diagram exemplifies the precision and clarity characteristic of professional crash documentation.

\begin{figure}[!htbp]
    \centering
    \begin{subfigure}[b]{0.45\textwidth}
        \centering
        \includegraphics[width=0.9\textwidth]{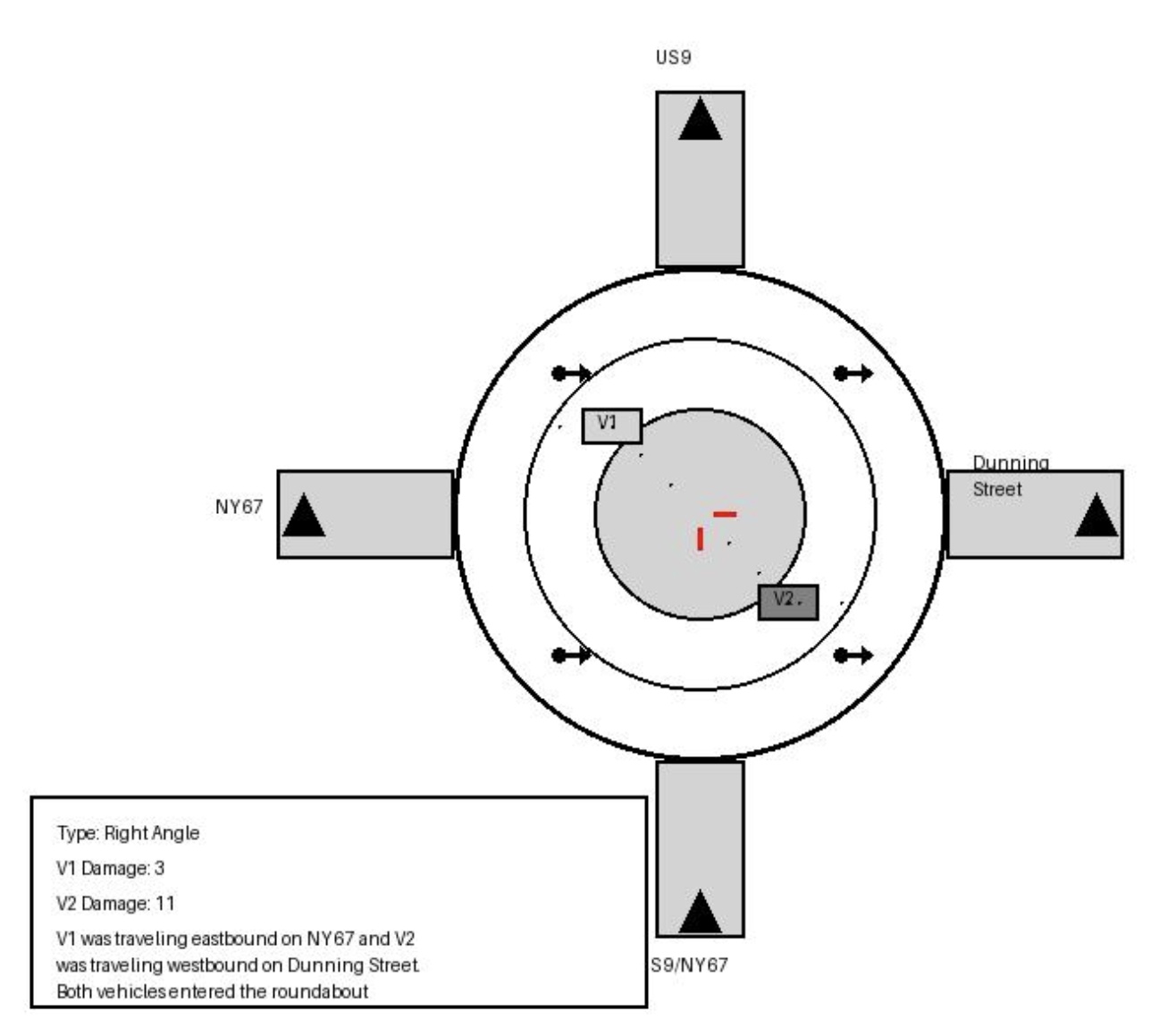}\vspace{-2mm}
        \subcaption{Generated crash diagram by Janus-4o}
        \label{fig:janus2}
    \end{subfigure}
    \hfill
    \begin{subfigure}[b]{0.45\textwidth}
        \centering
        \includegraphics[width=0.9\textwidth]{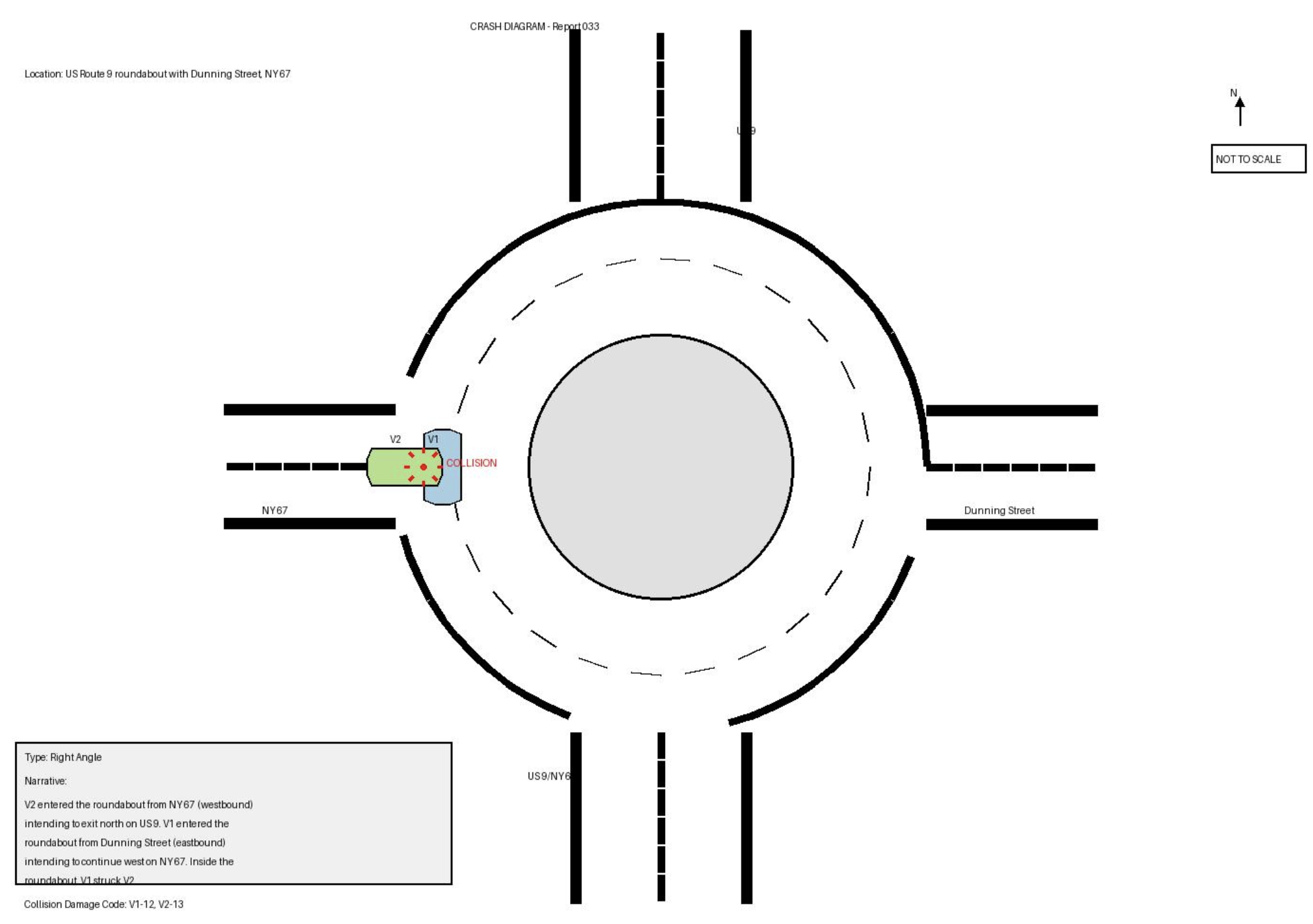}\vspace{-2mm}
        \subcaption{Generated crash diagram by Gemini-1.5-Flash}
        \label{fig:gemini2}
    \end{subfigure}
    \begin{subfigure}[b]{0.45\textwidth}
        \centering
        \includegraphics[width=0.75\textwidth]{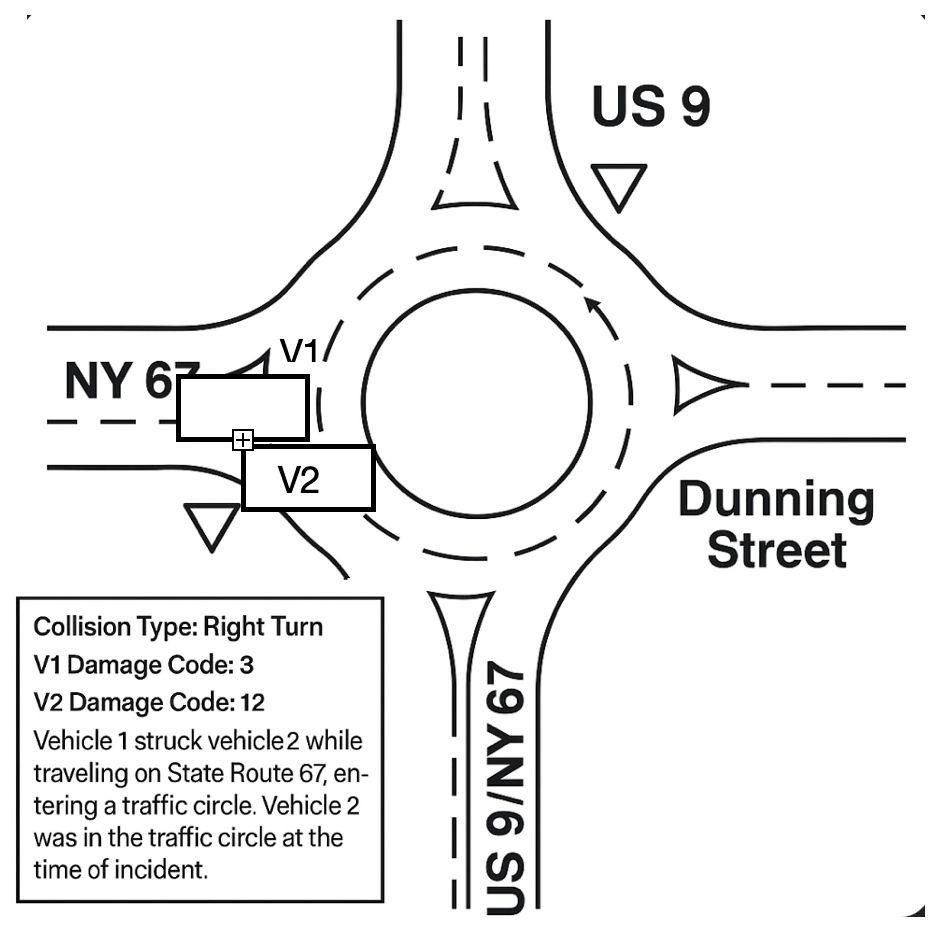}\vspace{-2mm}
        \subcaption{Generated crash diagram by GPT-4o}
        \label{fig:chatgpt2}
    \end{subfigure}
    \hfill
    \begin{subfigure}[b]{0.45\textwidth}
        \centering
        \includegraphics[width=0.9\textwidth]{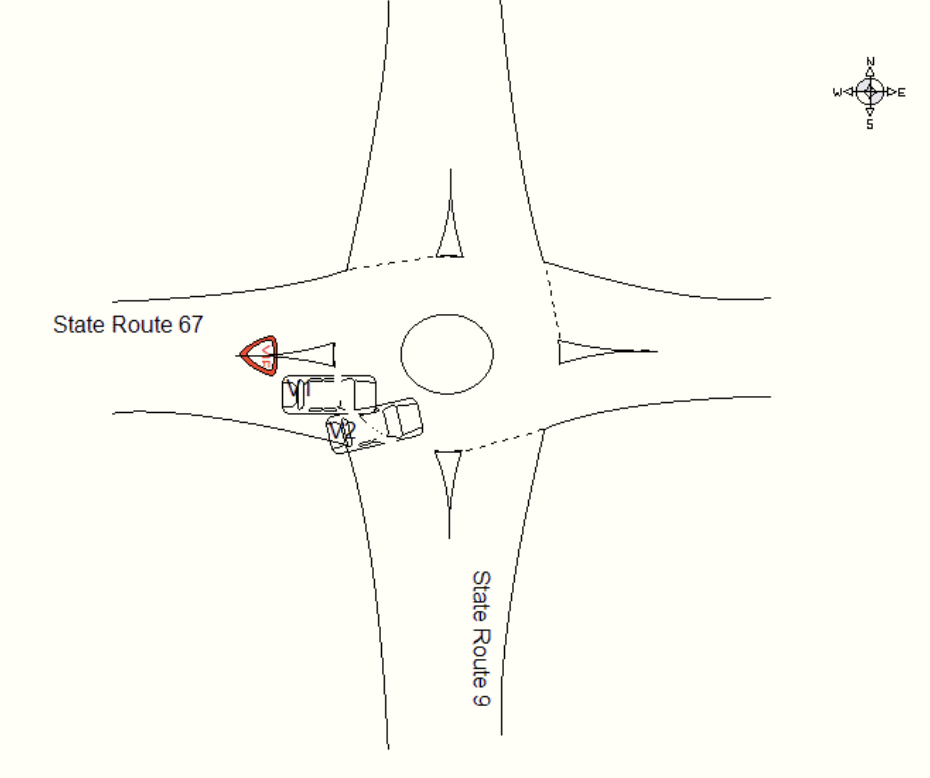}\vspace{-2mm}
        \subcaption{Crash diagram - ground truth (police report)}
        \label{fig:police2}
    \end{subfigure} 
    \caption{Comparison of crash diagrams generated by Janus{-4o} (top-left), Gemini-1.5-Flash (top-right), and GPT-4o (bottom-left) against the official police crash diagram (bottom-right) for the same crash.}
    \label{fig:comparison2}
\end{figure}

Overall, the visual comparison highlights differences in model capabilities. Janus-4o demonstrates notable structural and semantic deficiencies, such as the incorrect placement of vehicles and the omission of critical crash elements. Gemini-1.5-Flash offers improved spatial grounding and locational accuracy but encounters difficulties in fully representing vehicles and clearly depicting interactions. GPT-4o most effectively adheres to the structured roundabout format and prompt requirements, producing a coherent, well-labeled, and semantically comprehensive diagram that aligns with the anticipated analytical standards. Another example for a different crash is shown in Figure \ref{fig:comparison2}.

\section{Conclusions}

This study presents a systematic evaluation of three Vision-Language Models (VLMs): GPT-4o, Gemini-1.5-Flash, and Janus-4o, for crash diagram generation at a multilane roundabout. Beyond automating diagram sketching, the overarching objective of this work is to streamline the safety analysis workflow, enabling practitioners to aggregate AI-generated diagrams for pattern recognition, countermeasure development, and database standardization.

Among the models evaluated, GPT-4o demonstrated the strongest performance, particularly in accurately extracting semantic information from crash reports and generating  spatially coherent diagrams. It consistently produced outputs with correct collision type, accurate depiction of damage codes, clear vehicle labeling, and overall visual clarity. Nevertheless, GPT-4o occasionally misplaced collision points or misaligned directional arrows. In general, it outperformed Gemini-1.5-Flash and Janus-4o across most evaluation metrics. 

Gemini-1.5-Flash showed moderate strength in generating visually tidy diagrams and consistently labeling involved vehicles. However, its limitations became apparent when tasked with spatial reasoning or precise depiction of impact damage. In many cases, Gemini-1.5-Flash’s outputs lacked fidelity to the report's spatial context. In contrast, Janus-4o displayed the weakest alignment in its visual outputs. While visually clean and templated, Janus-4o generated diagrams frequently failed to reflect crash specifics, including collision types and point-of-impact locations.

These comparative results underscore the utility of VLMs for rapid, automated crash visualization while also exposing the need for continued refinement,particularly in spatial localization, geometric accuracy, and adherence to domain-specific rules. The structured prompt pipeline developed in this study offers a reproducible framework for future experiments and model benchmarking in the traffic safety domain. The proposed metric set for crash diagram evaluation demonstrates broad applicability across diverse crash scenarios, thereby providing a valuable methodological reference for future research in automating traffic safety visualization.

{Beyond fully automated diagram synthesis, a promising future direction is the development of hybrid pipelines in which a language model extracts structured crash information from free-text police reports, which is then passed to established crash diagram generation tools such as TIMS or Crash Magic. This approach could leverage the strengths of language models in semantic understanding and information extraction, while preserving the geometric fidelity, rendering quality, and workflow integration of professional-grade crash diagram software.}

{Future research will explore vector-based outputs, such as SVG formats compatible with GIS platforms, to improve editability, scalability, and integration with professional traffic-safety analysis workflows. Such outputs would also support downstream spatial analysis beyond static raster schematics.}

In conclusion, this study establishes a strong empirical foundation for the use of VLMs in automated crash diagram generation. It highlights the potential of advanced models like GPT-4o to significantly improve the efficiency and consistency of crash documentation processes, while also identifying key areas for future technical development. These insights are intended to guide researchers, practitioners, and policymakers in the responsible deployment of AI tools within the transportation safety domain.  With the rapid advancement of generative modeling, multimodal foundation models are expected to continue improving in capability and reliability. Future research should systematically evaluate emerging models and expand the dataset to encompass diverse intersection configurations, including signalized intersections, T-junctions, and mid-block crash scenarios. In addition, exploring vector-based outputs (e.g., SVG formats compatible with GIS platforms) would significantly enhance editability, interoperability, and seamless integration with existing traffic safety analysis tools. Such developments would strengthen real-world applicability while reducing reliance on static schematic representations.

\section{Limitations}

While the findings of this study are promising, several limitations warrant consideration. {The evaluation rubric used a binary (0/1) scoring scheme, which, while enabling consistent
comparison across models and raters, may limit granularity in assessing partial correctness.
Diagrams that approximately captured the correct quadrant, vehicle position, or damage
location but contained minor deviations were treated the same as entirely incorrect diagrams.
Future work will explore graded or weighted scoring schemes to better capture near-miss
spatial placements and partial damage localization.}

{The dataset consisted of 79 crash reports from a single two-lane roundabout in the Town of Malta, New York. This focused scope was a deliberate design choice to control for geometric variability and to evaluate model performance under a consistent and challenging intersection configuration. Nevertheless, this limits the generalizability of the findings to other intersection types, layouts, and jurisdictions. Future studies will expand the dataset to include signalized intersections, T-junctions, and mid-block crashes across diverse geographic contexts to assess model robustness and transferability.}

{The proposed framework relies on manually crafted structured prompts and fixed geometric templates to standardize inputs and enable controlled comparison across models. While this design improves experimental consistency, it may reduce robustness to variations in crash-report format, missing or ambiguous narrative information, and alternative roadway geometries. Future research will investigate adaptive prompting strategies and more flexible geometric representations to improve generalizability and transferability.} 

Although this study focused on crash diagram generation as a proxy for spatial-semantic understanding, it did not evaluate other potentially valuable outputs of VLMs, such as natural language crash summaries, violation classification, or injury risk prediction. Future research should explore these directions to assess the full spectrum of emerging VLMs in traffic safety applications.

{Another practical consideration not explicitly evaluated in this study is computational cost and inference time. VLMs differ substantially in latency, hardware requirements, and usage cost, which may influence their suitability for large-scale or real-time deployment. Future work will include systematic benchmarking of inference efficiency alongside diagram accuracy.}

{This study does not include comparisons with existing semi-automated or rule-based crash diagram generation methods, which typically rely on structured inputs or manual preprocessing rather than free-text police narratives. Including such baselines would help contextualize the practical advantages and limitations of VLMs. Future work will benchmark the proposed framework against representative tool-assisted and hybrid pipelines.} 
{Moreover, controlled robustness experiments, such as systematically varying
image scale, resolution, or layout were not conducted in this study. Evaluating sensitivity to such perturbations would provide deeper insight into model generalization.}

\bibliographystyle{unsrturl} 
\bibliography{references}

\end{document}